\documentclass[twocolumn,showpacs,superscriptaddress]{revtex4-1}
\usepackage{graphicx}
\usepackage{units}
\usepackage{multirow}
\usepackage{bigstrut}
\usepackage{overpic}
\usepackage{array}
\usepackage{color}
\usepackage{amsmath}
\newcommand{\dst}{D^{*}}
\newcommand{\dstbar}{\bar{D}^{*}}
\newcommand{\zcp}{Z^+_c(4025)}
\newcommand{\dstzerobar}{\bar{D}{}^{*0}}
\newcommand{\dstzero}{D^{*0}}
\newcommand{\dstplus}{D^{*+}}
\newcommand{\dstminus}{D^{*-}}
\newcommand{\dplus}{D^{+}}

\newcommand{\dzerobar}{\bar{D}{}^{0}}

\newcommand{\gm}{\gamma}

\newcommand{\ee}{e^+e^-}
\newcommand{\dstdstpi}{\dstplus\dstzerobar\pim}
\newcommand{\pip}{\pi^+}
\newcommand{\pizero}{\pi^0}
\newcommand{\pim}{\pi^-}
\newcommand{\jpsi}{J/\psi}
\newcommand{\psip}{\psi(2S)}
\newcommand{\mev}{\,\unit{MeV}}
\newcommand{\mevc}{\,\unit{MeV}/c}
\newcommand{\mevcc}{\,\unit{MeV}/c^2}
\newcommand{\gev}{\,\unit{GeV}}
\newcommand{\gevc}{\,\unit{GeV}/c}
\newcommand{\gevcc}{\,\unit{GeV}/c^2}

\begin{document}


\title{\boldmath Observation of a charged charmoniumlike structure in $e^+e^- \to (D^{*} \bar{D}^{*})^{\pm} \pi^\mp$  at $\sqrt{s}=4.26$\,GeV }

\author{
\small
\begin{center}
M.~Ablikim$^{1}$, M.~N.~Achasov$^{7,a}$, O.~Albayrak$^{4}$, D.~J.~Ambrose$^{40}$, F.~F.~An$^{1}$, Q.~An$^{41}$, J.~Z.~Bai$^{1}$, R.~Baldini Ferroli$^{18A}$, Y.~Ban$^{27}$, J.~Becker$^{3}$
% ...
\\
\vspace{0.2cm}
(BESIII Collaboration)\\
\vspace{0.2cm} {\it
$^{1}$ Institute of High Energy Physics, Beijing 100049, People's Republic of China\\
% ...
$^{3}$ Bochum Ruhr-University, D-44780 Bochum, Germany\\
$^{4}$ Carnegie Mellon University, Pittsburgh, Pennsylvania 15213, USA\\
% ...
$^{7}$ G.I. Budker Institute of Nuclear Physics SB RAS (BINP), Novosibirsk 630090, Russia\\
% ...
$^{18}$ (A)INFN Laboratori Nazionali di Frascati, I-00044, Frascati, Italy; (B)INFN and University of Perugia, I-06100, Perugia, Italy\\
% ...
$^{27}$ Peking University, Beijing 100871, People's Republic of China\\
% ...
$^{40}$ University of Rochester, Rochester, New York 14627, USA\\
$^{41}$ University of Science and Technology of China, Hefei 230026, People's Republic of China\\
% ...
$^{a}$ Also at the Novosibirsk State University, Novosibirsk, 630090, Russia\\
% ...
}\end{center}
\vspace{0.4cm}
}

\begin{abstract}

We study the process $e^+e^- \to (D^{*} \bar{D}^{*})^{\pm} \pi^\mp$ at
a center-of-mass energy of 4.26\,GeV using a 827\,pb$^{-1}$ data sample
obtained with the BESIII detector at the Beijing Electron Positron Collider.
Based on a partial reconstruction technique, the Born cross section is measured to be  $(137\pm9\pm15)$\,pb. We observe a structure near the $(D^{*} \bar{D}^{*})^{\pm}$ threshold in the $\pi^\mp$ recoil mass spectrum, which we denote as the $Z^{\pm}_c(4025)$.  The measured mass and width of the structure are $(4026.3\pm2.6\pm3.7)$\,MeV/$c^{2}$ and $(24.8\pm5.6\pm7.7)$\,MeV, respectively.  Its production ratio $\frac{\sigma(e^+e^-\to Z^{\pm}_c(4025)\pi^\mp \to (D^{*} \bar{D}^{*})^{\pm}
\pi^\mp)}{\sigma(e^+e^-\to (D^{*} \bar{D}^{*})^{\pm} \pi^\mp)}$ is determined to be $0.65\pm0.09\pm0.06$. The first uncertainties are statistical and the second are systematic.

\end{abstract}

\pacs{14.40.Rt, 13.66.Bc, 13.25.Gv}

\maketitle


Two charged bottomoniumlike particles, dubbed the $Z_b(10610)$ and
$Z_b(10650)$, have been observed in the $\pi^\pm \Upsilon(nS)$ and
$\pi^\pm h_b(mS)$ mass spectra at the Belle experiment in the decays
of $\Upsilon(10860)$ to $\pip\pim\Upsilon(nS) ~ (n=1,2,3)$ and to
$\pip\pim h_b(mP)~(m=1,2)$~\cite{Belle:2011aa}. Unlike a
conventional meson, the two states must involve at least four
constituent quarks to produce a non-zero electric charge. The masses of
the $Z_b(10610)$ and $Z_b(10650)$ are close to the
$B\bar{B}{}^*$ and $B^*\bar{B}{}^*$ thresholds,
respectively, which supports a molecular interpretation of $Z_b$'s as $B\bar{B}{}^*$ and
$B^*\bar{B}{}^*$ bound states~\cite{Bondar:2011ev}. In addition, this
scenario is supported by the subsequent observations of the decays
$Z_b(10610)\to B\bar{B}{}^*$ and $Z_b(10650)\to
B^*\bar{B}{}^*$ from the Belle experiment~\cite{Adachi:2012cx}.

A number of theoretical interpretations have been proposed to describe
the nature of the $Z_b$'s~\cite{Chen:2011cj,Yang:2011wz,Maiani:2004vq,Ali:2011ug}.
One intriguing suggestion is to look for corresponding particles in the charmonium sector~\cite{Yang:2011wz}.
As anticipated, a charged charmoniumlike structure, $Z_c(3900)$,  was
observed in the $\pi^\pm J/\psi$ mass spectrum in $\ee \to \pi^+\pi^- J/\psi$ by the BESIII experiment ~\cite{Ablikim:2013mio}, by the Belle experiment~\cite{Liu:2013dau} and using data from the CLEO-c experiment~\cite{Xiao:2013iha}.
More recently, BESIII has observed another charged state in the $\pi^\pm h_c$ mass
spectrum in $\ee \to \pi^+\pi^- h_c$, the $Z_c(4020)$~\cite{Ablikim:2013wzq}.
The masses of these states are slightly higher than the $D\dstbar$ and $\dst\dstbar$ mass
thresholds.
Therefore, a search of $Z_c$ candidates via their direct decays into
$\dst\dstbar$ pairs is strongly motivated.

In this Letter, we report on a study of the process $\ee\to
(\dst\dstbar)^\pm\pi^\mp$ at a center-of-mass energy
$\sqrt{s}=(4.260\pm0.001)\gev$, where  $(\dst\dstbar)^\pm$ refers to
the sum of the $\dstplus\dstzerobar$ and its charge conjugate
$\dstminus\dstzero$ final states. In the following, we use
the notation of  $\dstplus\dstzerobar$ and the inclusion of the charge
conjugate mode is always implied, unless explicitly stated. We use a partial reconstruction
technique to identify the  $\dstplus \dstzerobar\pim$ final states. This technique requires that only
the $\pim$ from the primary decay (denoted as the \emph{bachelor} $\pim$), the $\dplus$ decaying from $\dstplus \to
D^+ \pi^0$  and at least one soft $\pizero$ from $\dstplus \to D^+
\pi^0$ or $\dstzerobar \to \dzerobar \pi^0$ decay are reconstructed.
By reconstructing the $D^+$ particle,
the charges of its mother particle $\dstplus$ and the bachelor $\pim$
can be unambiguously identified.  Therefore, possible
combinatoric backgrounds are suppressed with respect to the signals.
We observe a charged charmoniumlike structure, denoted as
$\zcp$, in the $\pim$ recoil mass spectrum. The data presented in this Letter
correspond to an integrated luminosity of 827\,pb$^{-1}$,
which were accumulated with the BESIII detector~\cite{:2009vd} viewing $\ee$
collisions at the BEPCII collider~\cite{Zhang:2010zz}.


The BESIII detector is an approximately cylindrically symmetric detector with $93\%$
coverage of the solid angle around the $\ee$ collision point.
The apparatus relevant to this work includes, from inside to outside, a 43-layer
main wire drift chamber (MDC), a time-of-flight (TOF) system with two
layers in the barrel region and one layer for each end-cap,
and a 6240 cell CsI(Tl) crystal electro-magnetic calorimeter (EMC)
with both barrel and end-cap sections.
The barrel components reside within a superconducting solenoid
magnet providing a 1\,T magnetic field aligned with the beam axis.
The momentum resolution for charged tracks in the MDC
is $0.5\%$ for transverse momenta of $1\gevc$.
The energy resolution for showers in the EMC is $2.5\%$ for $1\gev$ photons.
For charged tracks, particle identification is accomplished by combining the measurements of the energy deposit registered in MDC, d$E$/d$x$,
and the flight time obtained from TOF to determine a probability $\mathcal{L}(h)~(h=\pi, K)$  for each hadron ($h$) hypothesis.
More details about the BESIII spectrometer are described elsewhere~\cite{:2009vd}.

Simulated data produced by the {\sc geant4}-based~\cite{Agostinelli:2002hh} Monte Carlo (MC) package, which
includes the geometric description of the BESIII detector and the
detector response, is used to optimize the event selection criteria,
to determine the detection efficiency and to estimate backgrounds.  The
simulation includes the beam energy spread and  initial-state
radiation (ISR)  modeled with {\sc
kkmc}~\cite{Jadach:2000ir}. The inclusive MC sample consists of
the production of the $Y(4260)$ state and its exclusive decays,
$\ee\to D^{(*)}\bar{D}^{(*)}(\pi)$, the production of ISR photons to
 low mass $\psi$ states and QED processes. Specific decays that are tabulated in
the Particle Data Group (PDG)~\cite{pdg2012} are modeled with
{\sc evtgen}~\cite{Lange:2001uf} and the unknown decay modes with
{\sc lundcharm}~\cite{Chen:2000tv}.  For the process
$\ee\to\dstdstpi$, ISR is included in the simulation, which
requires as input the cross section dependence on the center-of-mass energy.
For this, the observed cross sections for the process $\ee\to\dstdstpi$ at a sequence of energy values
around 4.260\,GeV at BESIII are used.
The maximum energy of the ISR photon in the simulation is $89\mev$,
corresponding to a $\dstdstpi$ mass
of $4.17\gevcc$.  For the resonant signal process
$\ee\to\zcp\pim\to\dstdstpi$, we assume that the $\zcp$ state has spin-parity of
$1^{+}$ and we simulate the cascade decays with angular distributions calculated from the corresponding matrix element. This assumption is consistent with our
observation in this analysis. However, other spin-parity assignments are not ruled out.


As discussed above, the reconstruction of the combinations of the $D^+$ and the
bachelor $\pim$ is used to identify
$\ee\to\dstdstpi$ final states. For the $D^+$ reconstruction, we only use
the $\dplus\to K^-\pip\pip$ decay, because it has dominant yields and the cleanest backgrounds compared to other $D^+$ deay modes.
We first select events
with at least four charged tracks. For each track, the polar angle
in the MDC must satisfy $|\cos\theta| < 0.93$ and the point of
closest approach to the $\ee$ interaction point must be within $\pm
10$\,cm in the beam direction and within 1\,cm in the plane
perpendicular to the beam direction. A $K(\pi)$ meson is identified by
requiring $\mathcal{L}(K)>\mathcal{L}(\pi)$
($\mathcal{L}(\pi)>\mathcal{L}(K)$). Among the identified tracks,
at least one $K^-$, two $\pip$'s and one $\pim$ are required in each event. For
the $\dplus\to K^-\pip\pip$ selection,  a vertex fit is implemented
that constrains the $K^-\pip\pip$ tracks to a common vertex; a fit
quality requirement is applied to suppress non-$\dplus$ decays.

\begin{figure}[tp!]
\centering
\begin{overpic}[width=0.48\linewidth]{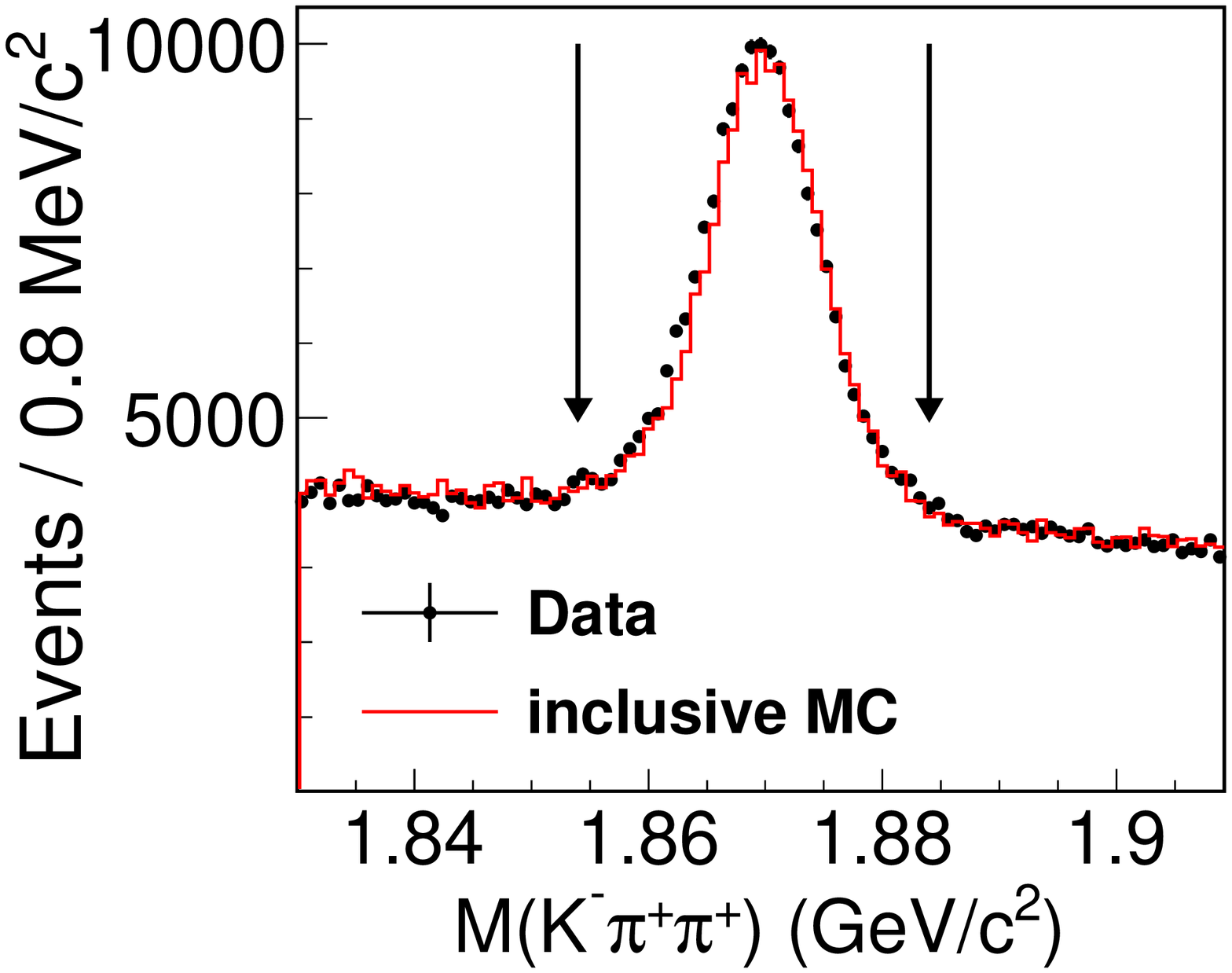}
\put(85,60){(a)}
\end{overpic}
\begin{overpic}[width=0.48\linewidth]{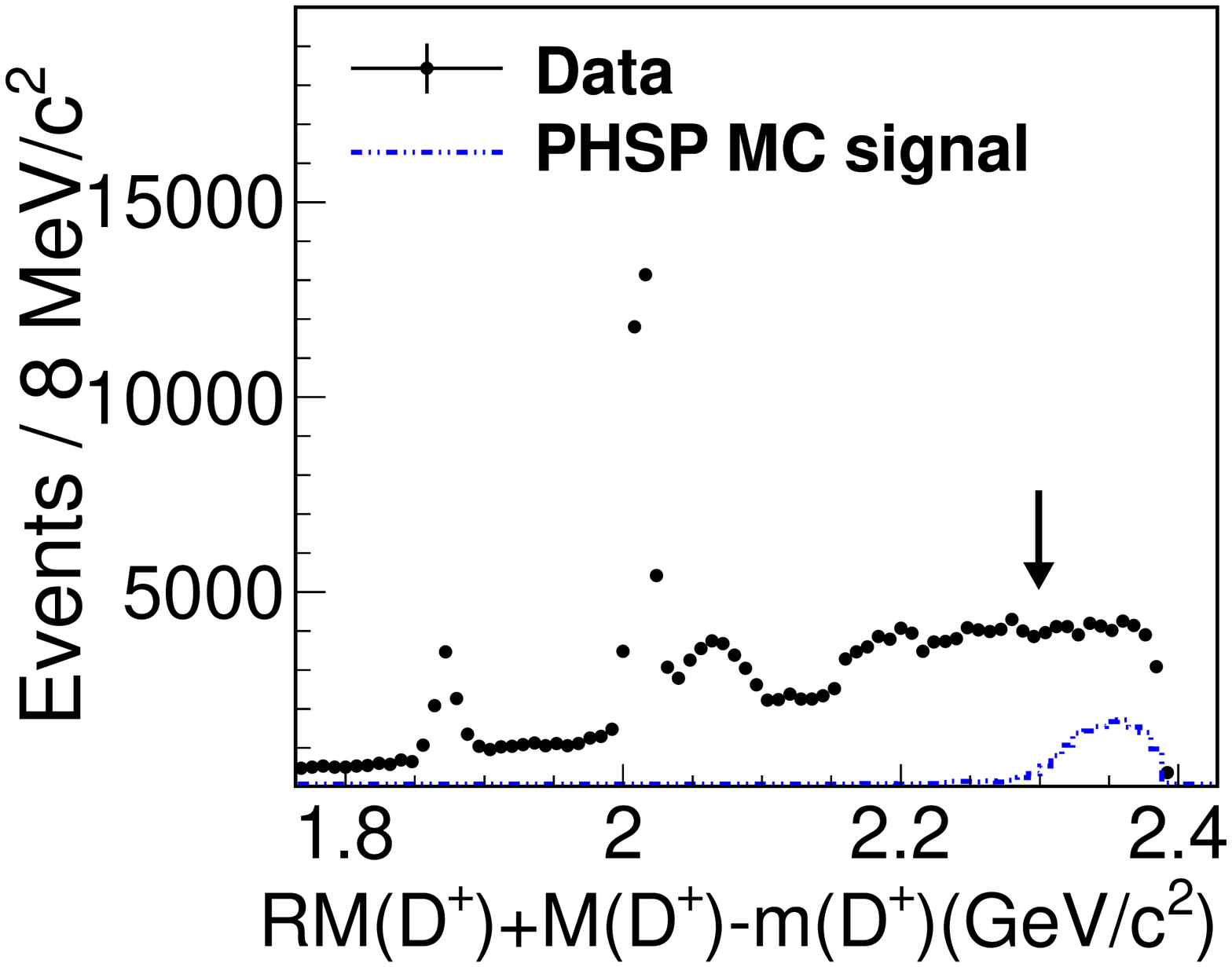}
\put(85,60){(b)}
\end{overpic}
\vspace{-0.5cm}
\caption{({\bf a}): a comparison of invariant mass $M(K^-\pip\pip)$ between data and MC simulation. The MC component is normalized to the area of the histogram of the data. Arrows indicate the mass region requirement. ({\bf b}): a comparison of $\dplus$ recoil mass  distributions between data and MC simulated three-body process $\ee\to\dstdstpi$ (PHSP signal). The level of the PHSP MC sample is scaled arbitrarily. The arrows show the position of the requirement $RM(\dplus)+M(\dplus)-m(\dplus)>2.3\gevcc$. See the text for a detailed description.}
\vspace{-0.5cm}
\label{fig:dmass}
\end{figure}

Figure~\ref{fig:dmass}(a) shows the $M(K^-\pi^+\pi^+)$ distribution where a $\dplus$ peak is clearly evident. All combinations with invariant mass in the region $(1.854, 1.884)\gevcc$ are identified as candidate $\dplus$ mesons.
The three peaks in the $\dplus$ recoil mass spectrum in Fig.~\ref{fig:dmass}(b) correspond, from left to right, to the two-body processes $\ee\to D^+D^-$, $D^+D^{*-}$ and $D^{*+}D^{*-}$, respectively. The $D^{*+}D^{*-}$ peak position corresponds to the sum of the $D^{*-}$ and $\pizero$ masses, since the soft $\pizero$ in $D^{*+}\to \dplus\pizero$ is missing.
The signal events lie at the rightmost side of the plot.
To improve the mass resolution, we exploit the correlations between $RM(\dplus)$ and $M(\dplus)$ and use $RM(\dplus)+M(\dplus)-m(\dplus)$  instead of $RM(\dplus)$. Here,  $RM(\dplus)$ is the recoil mass of the $\dplus$ candidate, $M(\dplus)$ is the
reconstructed mass of $\dplus$ candidate and $m(\dplus)$ is the world average $D^+$ mass~\cite{pdg2012}.
The recoil mass of $X$ is determined from
$RM(X)=|p_{\ee}-p_X|/c$, where $p_{\ee}$ and $p_X$ are the
four-momenta of the  initial $\ee$ systems and $X$ in the laboratory frame, respectively.
This technique is also used in
plotting other mass distributions presented in this paper.
Backgrounds from the two-body process $\ee\to D^{(*)}D^{(*)}$ are reduced by requiring
$RM(\dplus)+M(\dplus)-m(\dplus)>2.3\gevcc$.

\begin{figure}[t!]
\centering
\includegraphics[width=0.9\linewidth]{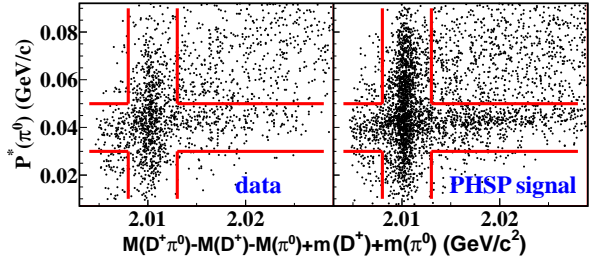}
\caption{Scatter plot of $P^*(\pizero)$ versus invariant mass of $D^+\pizero$ in data (left) and in PHSP signal MC (right).}
\vspace{-0.5cm}
\label{fig:tag_pi0}
\end{figure}

Additional background suppression is provided by requiring that
at least one $\pizero$ is reconstructed in the final
states. A $\pizero$ candidate is selected
by requiring at least two photon candidates
reconstructed from EMC showers~\cite{Ablikim:2012wwa} have an invariant mass in the range
$(0.120, 0.145)\gevcc$.
This $\pizero$ can be either from the
$\dstplus\to\dplus\pizero$ or $\dstzerobar\to\dzerobar\pizero$ decay.
In the case where the $\pizero$ is from $\dstplus\to\dplus\pizero$, the $\dplus\pizero$ invariant mass
peaks at the $\dstplus$ mass and a
mass region requirement
$2.008\gevcc<M(D^+\pi^0)-M(D^+)+m(D^+)-M(\pi^0)+m(\pi^0)<2.013\gevcc$
is used, corresponding to the vertical band in Fig.~\ref{fig:tag_pi0}.
In the case where the $\pizero$ is from $\dstzerobar\to\dzerobar\pizero$, its momentum
in the $\dplus\pim$  recoil system,
$P^*(\pizero)$, peaks at $43\mevc$
and a momentum requirement in the range  $(0.03, 0.05)\gevc$ is applied,
corresponding to the horizontal band in Fig.~\ref{fig:tag_pi0}. As
verified by MC simulations, the  $\dplus\pim$ recoil mass is
nearly the same as that of the  $\dstplus\pim$ recoil system, but is slightly
broadened due to the neglect of the soft $\pizero$ in the $\dstplus\to\dplus\pizero$ process.
Events with at least one $\pizero$ candidate, the one that fulfills either of the above requirements,
are retained.

\begin{figure}[t!]
\vspace{-0.25cm}
\centering
\includegraphics[width=\linewidth]{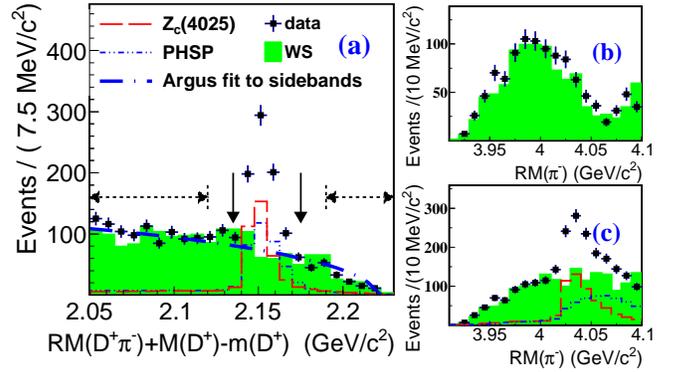}
\caption{ (\textbf{a}): spectra of recoil mass of $\dplus\pim$ with the exclusion of events, for which $RM(\pim)>4.1\gevcc$. Horizontal dotted-line arrows indicate the sidebands and vertical arrows indicate the signal region. The histogram of WS events is scaled by a factor of 1.9 to match the sideband data. (\textbf{b}) and (\textbf{c}): comparisons of  the $\pim$ recoil mass distributions between data and the WS events corresponding to the sideband and full regions as indicated in plot (a), respectively.}
\vspace{-0.5cm}
\label{fig:rmPiD}
\end{figure}

Figure~\ref{fig:rmPiD}(a) shows the $\dplus\pim$ recoil mass spectrum, where a
peak corresponding to the $\dstplus \dstzerobar\pim$ signal channel is evident.
The peak position roughly corresponds to the sum of the mass of $\dstzerobar$ and the mass of a $\pizero$, since the soft $\pizero$ that originates from the $\dstplus$ is not used in the computation of the recoil mass.
For other non-signal processes that have the same final state, such as $\ee\to
\dplus\pizero \dstzerobar \pim$, $\dstplus \dzerobar \pizero \pim$
and $\dplus \pizero\dzerobar\pizero\pim$, MC simulations of the phase space (PHSP) model do not produce narrow structures.
The distribution of combinatorial backgrounds is
estimated by combining a reconstructed $\dplus$ with a pion of the
wrong charge, referred to as  wrong-sign (WS) events.
The $\dplus\pim$ recoil mass distribution for the WS events, shown in Fig.~\ref{fig:rmPiD}(a),
is compatible with an ARGUS-function~\cite{Albrecht:1990am} shape fit to the sidebands of the signal peak in the data.
As shown in Fig.~\ref{fig:rmPiD}(b) and (c), the WS events with a scaling factor of 1.9 well represent the combinatorial backgrounds in the recoil mass spectra of the  bachelor $\pim$. This scaling is verified by an analysis of the inclusive MC data.
Backgrounds from the soft $\pim$  from $\dstminus$ decays in the $\ee\to\dstplus\dstminus(\pizero,\,\gm_{\unit{ISR}})$ processes are not well
described  by the WS background; its  $RM(\pim)$ distribution peaks in the region above $4.1\gevcc$, which is excluded in this analysis.

\begin{figure}[t!]
\centering
\includegraphics[width=0.8\linewidth]{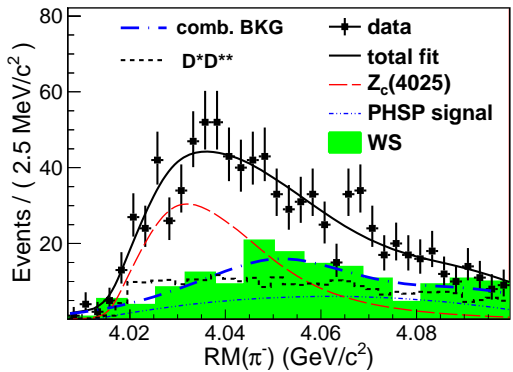}
\vspace{-0.25cm}
\caption{Unbinned maximum likelihood fit to the
   $\pim$ recoil mass spectrum in data. See the text for a detailed description of the various components that are used in the fit.  The scale of the $D^*D^{**}$ shape is arbitrary.}
\vspace{-0.25cm}
\label{fig:fit}
\end{figure}

In Fig.~\ref{fig:rmPiD}(c), a clear enhancement above the WS background is evident.
To study the enhancement, the events of the $\dstplus \dstzerobar\pim$ final states within the signal region $(2.135, 2.175)\gevcc$ in Fig.~\ref{fig:rmPiD}(a) are selected and displayed in Fig.~\ref{fig:fit}.
The enhancement cannot be attributed to the PHSP $\ee\to \dstplus \dstzerobar\pim$
process. We simulate the processes of $\ee\to
D^{**}\bar{D}^{(*)}, D^{**}\to D^{(*)}\pi(\pi)$, where $D^{**}$ denotes
neutral and charged highly excited $D$ states, such as $D_0^*(2400)$,
$D_1(2420)$, $D_1(2430)$ and $D_2^*(2460)$.  Among these processes,
only those with $\dstplus\dstzerobar\pim$  final states, which are
not components of the WS backgrounds, would contribute to the
difference between data and the WS backgrounds.
No peaking structure in the $\pim$ recoil mass spectra for these simulated events is seen in Fig.~\ref{fig:fit}.
Since the energy $\sqrt{s}=4.26\gev$ is much lower than the production thresholds of
$D^{**}\bar{D}^{*}$, we neglect the possibility of backgrounds relevant to $D^{**}\bar{D}^{*}$ processes.

The observed enhancement is very close to the $m(\dstplus)+m(\dstzerobar)$
mass threshold. We assume that  the enhancement is due to a particle, labeled as
$\zcp$, and parameterize its line shape by the product of an $S$-wave  Breit-Wigner
(BW) shape and a phase space factor $p\cdot q$
\begin{equation}
 \label{eq:bw}
  \Big | \frac{1}{M^2-m^2 + i m\Gamma/c^2} \Big | ^2\cdot p\cdot q .
\end{equation}
Here, $M$ is the reconstructed mass; $m$ is the resonance mass; $\Gamma$ is the width; $p(q)$ is
the $\dstplus(\pim)$ momentum in the rest frame of the
$\dstplus\dstzerobar$ system (the initial $\ee$ system).

The signal yield of the $\zcp$ is estimated by an unbinned maximum
likelihood fit to the spectrum of $RM(\pim)$. The fit results are
shown in Fig.~\ref{fig:fit}. Possible interference between the $\zcp$
signals and the PHSP processes is neglected. The $\zcp$ signal shape
is taken as an efficiency-weighted BW shape convoluted with a detector
resolution function, which is obtained from MC simulation. The detector resolution is about $2\mevcc$ and
is asymmetric due to the  effects of ISR. The shape of the combinatorial
backgrounds is taken from the kernel-estimate~\cite{Cranmer:2000du} of the WS events and its magnitude is
fixed to the number of the fitted background events within the
signal window in Fig.~\ref{fig:rmPiD}(a). The shape of the PHSP signal
is taken from the MC simulation and its amplitude is taken as a free parameter in the
fit. By using the MC shape, the smearing due to  effects of ISR and the
detector resolution are taken into account. From the fit, the parameters of $m$ and $\Gamma$ in Eq.~\eqref{eq:bw}
are determined to be
\begin{eqnarray}
   m(\zcp)&=&(4026.3\pm2.6)  \mevcc, \nonumber\\
  \Gamma(\zcp)&=&(24.8 \pm 5.6) \mev.   \nonumber
\end{eqnarray}
A goodness-of-fit test gives a $\chi^2/$d.o.f.$=30.4/33=0.92$.
The $\zcp$ signal is observed with a statistical significance of
13\,$\sigma$, as determined by the ratio of the maximum likelihood
value and the likelihood value for a fit with a null-signal
hypothesis. When the systematic uncertainties are taken into
account, the significance is evaluated to be 10\,$\sigma$.

The Born cross section is determined from
$\sigma=\frac{n_\mathrm{sig}}{\mathcal{L}(1+\delta)\varepsilon\mathcal{B}}$,
where $n_\mathrm{sig}$ is the number of  observed signal events,
$\mathcal{L}$ is the integrated luminosity, $\varepsilon$ is the
detection efficiency, $1+\delta$ is the radiative correction factor
and $\mathcal{B}$ is the branching fraction of
${\dstplus}\to\dplus(\pizero, \gm)$, $\dplus\to K^-\pi^+\pi^+$. From the
fit results, we obtain
$560.1\pm30.6$ $\dstplus \dstzerobar\pim$ events,
among which $400.9\pm47.3$ events are $\zcp$ candidates. With the input
of the observed center-of-mass energy dependence of $\sigma(\dstplus \dstzerobar\pim)$,
the radiative correction factor is calculated
to second-order in QED~\cite{Kuraev:1985hb} to be $0.78\pm0.03$. The efficiency for the
$\zcp$ signal process is determined to be 23.5\%, while the efficiency
of the PHSP signal process is 17.4\%.  The total cross section
$\sigma(\ee\to (\dst{\dstbar})^\mp\pi^\pm)$ is measured to be
$(137\pm9)\,\unit{pb}$,  and the ratio $R=\frac{\sigma(\ee\to
Z^{\pm}_c(4025)\pi^\mp \to (D^{*} \bar{D}^{*})^{\pm}
\pi^\mp)}{\sigma(\ee\to (D^{*} \bar{D}^{*})^{\pm} \pi^\mp)}$ is determined to be $0.65\pm0.09$.

\begin{table}[t]
  \begin{center}
  \begin{tabular}{l|ccccc}
      \hline \hline
      Source   & $m$($\unit{MeV}/c^{2}$)  & $\Gamma$($\unit{MeV}$)  &  $\sigma_\textrm{tot}$(\%) &  $R$(\%)  \\ \hline
      Tracking  &  & & 4 \\
      Particle ID &  &  & 5 \\
      Tagging $\pizero$  & & & 4  \\
      Mass  scale  &   1.8 &  &  &  \\
      Signal shape  &  1.4 & 7.3 & 1 & $5$   \\
      Backgrounds  &  1.5 & 0.6 & 5 & $5$  \\
      Efficiencies  &  0.9 & 2.2 & 1 & 5  \\
      $D^{**}$ states & 2.2  & 0.7 & 5 & $2$ \\
      Fit range  & 0.9 & 0.9 & 1 & 1   \\
      $\dstplus \dstzerobar\pim$ line shape & & & $4$ &  \\
      PHSP model & & & $2$ & 2 \\
      Luminosity &  & & 1.0  & \\
      Branching fractions &  & & 2.6 &   \\
          \hline
          total & 3.7 & 7.7 & 11  & 9   \\
        \hline\hline
    \end{tabular}
    \caption{A summary of the systematic uncertainties on the measurements of the $\zcp$ resonance parameters and cross sections.
    We denote $\sigma_\textrm{tot}=\sigma(\ee\to (\dst\dstbar)^\pm\pi^{\mp})$.
    The total systematic uncertainty is
    taken as the square root of the quadratic sum of the individual uncertainties. }
    \label{tab:sys_err}
   \vspace{-0.6cm}
  \end{center}
  \end{table}

Sources of systematic error on the measurement of the $\zcp$ resonance
parameters and the cross section are listed in Table~\ref{tab:sys_err}.
The main sources of systematic uncertainties relevant for determining the $\zcp$ resonance parameters
and the ratio $R$ include the mass scale, the signal shape, background
models and potential $D^{**}$ backgrounds. We use the process
$\ee\to\dplus\dstzerobar\pim$ to study the mass scale of the
recoil mass of the low momentum bachelor $\pim$. By fitting
the peak of $\dstzerobar$ in the $\dplus\pim$ recoil mass spectrum,
we obtain a mass of $2008.6\pm0.1\mevcc$. This
deviates from the PDG reference value by $1.6\pm0.2\mevcc$. Since the fitted variable
$RM(\dplus\pim)+M(\dplus)-m(\dplus)$ removes the correlation with
$M(\dplus)$, the shift mostly is due to the momentum measurement of the bachelor $\pim$. Hence,
we take the mass shift of $1.8\mevcc$ as a systematic uncertainty on $RM(\pim)$ due
to the mass scale.
If one assumes $\zcp$ also decays to other
final states such as $\pip(\psip, \jpsi, h_c)$, variations of their
relative coupling strengths would affect the measurements of the $\zcp$
mass and width. The Flatt\'{e} formula~\cite{flatte} is used to take into account possible multiple channels,
and the maximum
changes on the mass and the width are $0.4\mevcc$ and $0.1\mev$,
respectively. When we assume that the relative momentum between the $\pim$
and $\zcp$ in the rest frame of the $\ee$ system is a $P$-wave, the mass
and width change from the nominal results by $1.4\mevcc$ and
$7.3\mev$, respectively. The maximum variations are taken as systematic
uncertainties. Variations in the unbinned and non-parametric
kernel-estimate of the WS events and
fluctuations of the estimated numbers of combinatorial backgrounds
give maximum changes of $1.5\mevcc$ in the mass, $0.6\mev$ in the width, 5\%
in the total cross section and 5\% in the ratio $R$. We vary the
parameters of the BW shape used to model the $\zcp$ signals in the MC
simulation; the mass is changed in the range of $(4.02, 4.04)\gevcc$ and the
width is changed in the range of $(20, 45)\mev$. All these
variations would influence the efficiency curves and thereby, affect
the cross section results. The maximum changes are taken into
account as systematic uncertainties. We performed a fit with  the inclusion
of the possible backgrounds due to the $\ee\to D^{**}D^*$ processes
in the $RM(\pim)$ spectrum. The resultant changes are taken as a
systematic uncertainty.

The spin-dependence of the non-resonant process is
studied by changing the orientation of the decay plane and by changing the relative angular distributions among the final states of $\dstplus\dstzerobar\pim$. The influences on the measurements of the cross section and the ratio $R$ are at the 2\% level.
Other items in Table~\ref{tab:sys_err} mostly influence the measurement
of the total cross section.
The efficiencies of the soft $\pi^\pm$ are well understood in MC simulation~\cite{Ablikim:2013pqa}.
Uncertainties associated with the
efficiencies of the tracking and the identification of the four final charged
track are estimated to be 4\% and 5\%, respectively.  A possible bias in the efficiency determination
for tagging the $\pizero$ is estimated to be 4\% by
comparing the measurements of $\sigma(\ee\to \dstplus
\dstzerobar\pim)$ with and without detecting the $\pizero$. The  line shape of the $
\dstplus \dstzerobar\pim$ cross sections affects the
radiative correction factor and the detection efficiency simultaneously.
This uncertainty is estimated to be 4\% by changing the input of the
observed line-shape within errors.
The uncertainty of the integrated
luminosity, measured with large angle Bhabha events, is determined
to be 1\%. Branching fractions for $\dstplus\to\dplus(\pizero, \gm),
\dplus\to K^-\pi^+\pi^+$ are used in calculating the cross section
and their uncertainty taken from the PDG~\cite{pdg2012} is included
as a systematic uncertainty.

To summarize, we observe an enhancement near
the threshold of $m(\dstplus)+m(\dstzerobar)$ in the $\pi^\mp$ recoil
mass spectrum in the process $\ee \to (\dst\dstbar)^{\pm}\pi^\mp$ at
$\sqrt{s}=4.260\gev$. If the enhancement is due to a charmoniumlike particle,
namely $Z^\pm_c(4025)$, its mass and width are measured to be
$(4026.3\pm2.6 \pm 3.7) \mevcc$ and $(24.8 \pm 5.6 \pm 7.7)\mev$,
respectively. 
To validate the establishment of the $Z_c(4025)$, a rigorous spin analysis is required based on a larger data sample.
Since the $Z_c(4025)$ couples to $(\dst\dstbar)^{\pm}$ and has electric charge, the observation suggests that the $Z_c(4025)$ may be a
virtual $\dst\dstbar$ resonant system~\cite{Yang:2011wz}.
The resonance parameters of the $Z_c(4025)$ agree with the $Z_c(4020)$ within $1.5\,\sigma$~\cite{Ablikim:2013wzq}. To identify whether they are same particle, one needs a further sophisticated analysis with a coupled channel technique.
The Born cross section $\sigma(\ee\to (\dst\dstbar)^\pm\pi^{\mp})$ is measured to be
$(137\pm9\pm15)$\,pb, based on a second-order QED calculation, which
is compatible with CLEO-c's result~\cite{CroninHennessy:2008yi},
assuming that isospin symmetry is not largely broken. The first uncertainties are
statistical and the second are systematic.
The ratio $R=\frac{\sigma(\ee\to Z^\pm_c(4025)\pi^\mp \to (\dst
\dstbar)^\pm\pi^{\mp})}{\sigma(\ee\to (\dst\dstbar)^\pm\pi^{\mp})}$ is
determined to be $0.65\pm0.09\pm0.06$. 


The BESIII collaboration thanks the staff of BEPCII and the computing center for their strong support. This work is supported in part by the Ministry of Science and Technology of China under Contract No. 2009CB825200; Joint Funds of the National Natural Science Foundation of China under Contracts Nos. 11079008, 11179007, U1332201; National Natural Science Foundation of China (NSFC) under Contracts Nos. 10625524, 10821063, 10825524, 10835001, 10935007, 11125525, 11235011, 1127526; the Chinese Academy of Sciences (CAS) Large-Scale Scientific Facility Program; CAS under Contracts Nos. KJCX2-YW-N29, KJCX2-YW-N45; 100 Talents Program of CAS; German Research Foundation DFG under Contract No. Collaborative Research Center CRC-1044; Istituto Nazionale di Fisica Nucleare, Italy; Ministry of Development of Turkey under Contract No. DPT2006K-120470; U. S. Department of Energy under Contracts Nos. DE-FG02-04ER41291, DE-FG02-05ER41374, DE-FG02-94ER40823, DESC0010118; U.S. National Science Foundation; University of Groningen (RuG) and the Helmholtzzentrum fuer Schwerionenforschung GmbH (GSI), Darmstadt; WCU Program of National Research Foundation of Korea under Contract No. R32-2008-000-10155-0.


\end{document}